\documentclass[twocolumn,aps,pra,showpacs]{revtex4}
\usepackage{epsfig} \begin{document} \title{Quantum adaptation of
noisy channels} \author{Miroslav Gavenda and Radim Filip}
\affiliation{Department of Optics, Palack\' y University, 17.
listopadu 50,  772~07 Olomouc, Czech Republic} \date{\today}
\begin{abstract}
Probabilistic quantum filtering is proposed to
  properly  adapt sequential independent quantum channels in order to stop
  sudden death of entanglement. In the adaptation, the quantum
  filtering does not distill or purify more entanglement, it rather
  properly prepares entangled state to the subsequent quantum channel. For
  example, the quantum adaptation probabilistically eliminates the sudden
  death of entanglement of two-qubit entangled state with isotropic
  noise injected into separate amplitude damping channels. The result
  has a direct application in quantum key distribution through noisy
  channels.
\end{abstract}  \maketitle

\section{Introduction}

Quantum entanglement propagating through noisy quantum channel is
crucial for modern application of quantum information science, for
example, for security of quantum key distribution \cite{qc} or
cluster-state quantum computing \cite{qcomp}. The quantum channel
represents a real physical system used to transmit, store or operate
quantum states. Recently, it has been recognized that the entanglement
can be even lost if a noisy entangled state is inserted into a channel
which is entanglement preserving for all the channel parameters (such
as the amplitude and phase damping channel)
\cite{Yu04,Yu06,Yu07,Davidovitch07}. Such the behavior is called the
sudden death of entanglement and it can dramatically reduce the
security of the
key distribution or the efficiency of cluster-state quantum computing.
Therefore, it is interesting how to stop the sudden death of
entanglement, deterministically or even probabilistically, at a cost
of success rate of the transmission of entanglement.

The sudden death of entanglement has been reported as a property of a
given non-maximally entangled state passing through the quantum channel 
representing a finite-time continuous interaction with a reservoir. A state 
preparation of the non-maximally entangled state is considered to be independent 
from the subsequent noisy channel. Let us to describe this situation in more 
abstract way.
Such the non-maximally entangled state can be generally
understand as an output from some previous independent channel applied on this 
maximally entangled state. 
Thus we have a composite of two independent channels. The reservoirs 
corresponding to these two channels are therefore considered to be independent.
It very well corresponds to a broad class of realistic physical situations in 
which the reservoirs of two channels are not interacting. Optimizing
over input represented by 
maximally entangled states it can be recognized whether the composite
channel,
exhibiting sudden death of entanglement, is actually entanglement
breaking channel \cite{Ruskai03}. For the entanglement breaking
channel, no entanglement propagates through the channel. 
Also no entanglement can be 
distilled after the channel.
Below, the sudden death of entanglement will be understand
rather as a property of a composition of channels with independent
reservoirs. It is not always possible to split a given quantum channel
into independent sub-channels. Then the reservoirs corresponding to sub-channels 
are not independent and their exact dynamics and coupling have to be studied in a 
detail. We will focus just on the case of independent channels.

We study two-qubit entanglement undergoing local unitary quantum
dynamics. This means that each qubit interacts just with its
own reservoir resulting in channels
for which the Kraus decomposition is valid
\cite{Hayashi}. We concern only
about the cases when the sudden death of entanglement in the composite
channel breaks entanglement completely. To stop the sudden death of
entanglement, single-copy distillation \cite{Verstraete01,Wang06} can
be sometimes simply placed between the channels. Then distillation increases
entanglement before the subsequent channel and
any construction of the distillation is only optimized with a respect
to the state after the previous channel.

In this paper, the adaptation of quantum channels is proposed to
prevent the sudden death of entanglement. The probabilistic adaptation
differs from the single-copy entanglement distillation and
purification \cite{Verstraete01,Wang06}. In the adaptation,
even in the case that the
entanglement could not be increased by single-copy distillation after first
channel, still the entangled state can be better
prepared to the subsequent channel to preserve
entanglement. Basically, the proper adaptation depends on the
subsequent noisy channel. As will be demonstrated, it can help to stop
the sudden death of entanglement when the
single-copy distillation is inefficient. It is rather complex problem to find
generally optimal adaptation between the channels. Therefore we rather
discuss realistic examples of the sudden death of entanglement to
demonstrate a potential power of the adaptation. First, we concentrate
on a simple example of two subsequent single-qubit non-unital channels
with an anisotropic noise. We show by optimal unitary
adaptation of such the channels, the sudden death of entanglement is
completely canceled for all the channel parameters. Second,
probabilistic adaptation between the channel with an isotropic noise
and the subsequent amplitude damping channel is analyzed. In this
case, the sudden death of entanglement cannot be stopped by the
unitary adaptation. But if the probabilistic adaptation is applied
then the sudden death of entanglement vanishes completely. It
demonstrates, that the sudden death of entanglement can be partially or
even fully caused by the improper adaptation of the noisy channels. To
find whether the sudden death of entanglement is really presented, it
is necessary to discuss the optimal adaptation and then analyze if the
entanglement passing through the adapted channels will be broken or
not. Such the results have a direct application in quantum key
distribution through composite realistic channels and in a multi-qubit version,
also in the
cluster-state preparation for quantum computing.

In Sec. II we introduce our method of quantum adaptation for
composition of
independent quantum channels. We use the Kraus decomposition
for the representation of the channels and give arguments for its
validity.
We accent two different configurations of quantum channels, asymmetrical and
symmetrical one and explain what is a non-trivial feature of the
sudden death of entanglement. We introduce the single-copy quantum filtering
operations between the channels to help to stop the sudden death of
entanglement.
In Sec. III
 we analyze simple example of quantum
adaptation for the asymmetrical
configuration of simple quantum channels. For this case the sudden
death of entanglement can be undone by simply performing an appropriate
unitary transform between the channels.
In Sec. IV we show that in symmetrical configuration of quantum
channels a unitary transform is not enough to undone the break of
entanglement.
Introducing quantum filters between the
channels can help to stop the entanglement breaking for the symmetrical
configuration.
We conclude in Sec. V.

\section{Quantum Adaptation of Independent channels}
From our point of view, the sudden death of entanglement is an
entanglement breaking property of a composition of independent quantum channels
if at least single one is not entanglement breaking at all.
We assume any particle from entangled state evolves locally unitarily
and reacts just with its own reservoir. Reservoirs are mutually
independent. For such a case we may
describe the evolution of the entangled state by using
independent quantum channels in the form of Kraus
decomposition \cite{Kraus}
\begin{equation}
  \chi(\rho)=\sum_{i=1}^{N}A_i\rho A_i^{\dag},
\end{equation} where
$\sum_{k}A_k^{\dag}A_k=\openone$. An important class of the channels are
unital channels (for example, depolarization channel or
phase damping channel), which preserves the isotropic noise. Such
the channels transform maximal entangled state only to a mixture
of maximally entangled states. The channel is entanglement breaking
if no entanglement remains although any maximally entangled state
$|\Psi\rangle_{AB}$ is passing through the channel \cite{Ruskai03}.
Mathematically, for single-qubit channel it corresponds to a
condition based on positive partial transposition criterion
\cite{PPT}, explicitly $[\chi_B(\rho_{AB})]^{T_A}\geq 0$, where
$\rho_{AB}=|\Psi\rangle_{AB}\langle\Psi|$. After such the
channel no entanglement can be distilled even by multi-copy
distillation \cite{Bennett96}.

The basic asymmetrical and symmetrical configurations of the propagation of
two-qubit maximally entangled state through independent noisy
channels are depicted on Fig.~1. In the first case, a single qubit
from maximally entangled two-qubit state
$\rho_{AB}=|\Psi\rangle_{AB}\langle\Psi|$ is propagating through two
independent channels $\chi_{B1}$ and $\chi_{B2}$ having mutually
independent reservoirs. In the second case,
both the qubits are symmetrically propagating through independent (but
simply identical) consecutive channels $\chi_{A1},\chi_{A2}$ and
$\chi_{B1},\chi_{B2}$ all having mutually independent reservoirs.
The maximally entangled state $\rho_{AB}$
passing through the asymmetrical composition of two channels
$\chi_{B2}\circ\chi_{B1}(\rho_{AB})$ can be described as
\begin{equation}
\chi_{B2}\circ\chi_{B1}(\rho_{AB})=\sum_{i=1}^{N_1}
\sum_{j=1}^{N_2}B_{2j}B_{1i}\rho_{AB}
  B_{1i}^{\dag}B_{2j}^{\dag}
\end{equation} and the composite channel
is entanglement breaking if
\begin{equation}\label{cond1}
    [\chi_{B2}\circ\chi_{B1}(\rho_{AB})]^{T_A}\geq
    0.
\end{equation}
For the symmetrical configuration, the maximally
entangled state going through the symmetrical both-side
channels is transformed to
\begin{eqnarray}
        \chi_{A2}\circ\chi_{A1}\circ\chi_{B2}\circ\chi_{B1}(\rho_{AB})=\nonumber\\
         \sum_{i,k=1}^{N_1}\sum_{j,l=1}^{N_2}
         B_{2i}B_{1j}A_{2k}A_{1l}\rho_{AB}
         A_{1l}^{\dag}A_{2k}^{\dag}B_{1j}^{\dag}B_{2i}^{\dag}
\end{eqnarray}
and the composite channel is entanglement
breaking if
\begin{equation}\label{cond2}
         [\chi_{A2}\circ\chi_{A1}\chi_{B2}\circ\chi_{B1}(\rho_{AB})]^{T_A}\geq
         0.
\end{equation}
For the asymmetrical situation, the
entanglement breaking property does not depend on a kind of
maximally entangled state \cite{Ruskai03}, but for the
symmetrical configuration it is not generally true
\cite{Yu07}. The channel is sequentially entanglement
preserving if
\begin{equation}\label{cond3}
           [\chi_{Bi}(\rho_{AB})]^{T_A}< 0,\,\,\,
           [\chi_{Ai}(\rho_{AB})]^{T_A}< 0
\end{equation}
and only
such the channels will be taken into consideration. If the
channel is not sequentially entanglement preserving then
the entanglement cannot be successfully transmitted with a
help of any adaptation method. Also the composite channels
preserving entanglement will be simply omitted in the
following discussion. It is obvious that two consecutive
channels can break entanglement although they do not break
it separately.
But this is not non-trivial effect of the sudden death of
entanglement. A non-trivial feature of the ESD
\cite{Yu04,Yu06,Yu07,Davidovitch07} is
that each channel is entanglement preserving separately but the
whole concatenation (whole channel) is
entanglement breaking. 
Physically, we have different independent channels preserving entanglement for 
any value of finite channel parameters. 
But their combination exhibiting the sudden death 
of entanglement can give entanglement breaking channel for some
values of channels parameters. Since it is impossible improve 
the composite channel just by the operations before and 
after the channel, it is non-trivial case of the ESD. 

\begin{figure}
\centerline{\psfig{width=7cm,angle=0,file=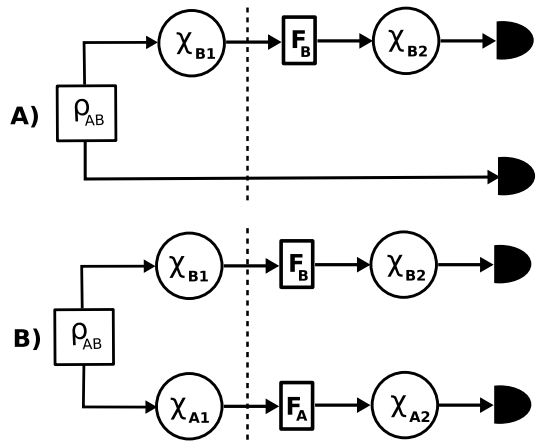}}
\caption{The channel adaptation for the maximally entangled state passing through
asymmetrical (A) and symmetrical (B) pairs of the independent channels
with local unitary dynamics, dashed lines part the protocol virtually
to the preparation stage (mixed state from maximally entangled state)
(left) and to the other stages (right):
$\rho_{AB}$ -- maximally entangled state, $\chi_{Ai,Bi}$ -- quantum
channels, $F_{A,B}$ -- quantum filters.} \end{figure}

To stop the sudden death of entanglement in the composite channel, a
single-copy quantum filter can be generally placed between the
individual channels. The quantum filter on single-qubit state can be
described by the transformation
\begin{equation}
\rho'=\frac{F\rho
  F^{\dag}}{\mbox{Tr}(F\rho F^{\dag})},
  \end{equation}
where
$F^{\dag}F\leq \openone$. Generally, the quantum filtering can be
decomposed as $F=UF_0V$, where $U,V$ are single-qubit unitary
operations and $F_0=\mbox{diag}(1,\sqrt{r})$, $0\leq r\leq 1$. If $r=1$
the filtering is reduced just to unitary operation. It was
theoretically described that such the local filtration applied after
the noisy channel can increase entanglement \cite{Verstraete01}.
Recently, the local filtering has been experimentally demonstrated
for entangled pair of photons generated from spontaneous parametric
down-conversion \cite{Wang06}. The result of the optimal local
filtration always approaches the mixture of Bell diagonal states,
from which more entanglement cannot be further distilled by any
single-copy local filters. It automatically excludes a chance to stop the
entanglement sudden death for a composition of the unital channels.
On the other hand, for the non-unital channels the single copy
filtration could be able to increase entanglement or perform
proper adaptation of the channels.

With the filters between the channels, after the
asymmetrical channels in the configuration (A) the maximally entangled state is
\begin{eqnarray}
\chi_{B2}\circ F_B\circ
\chi_{B1}(\rho_{AB})=\nonumber \\
\frac{1}{S}\sum_{i=1}^{N_1}
\sum_{j=1}^{N_2}B_{2j}F_BB_{1i}\rho_{AB}
  B_{1i}^{\dag}F_B^{\dag}B_{2j}^{\dag},
\end{eqnarray}
where the success rate is
$S=\sum_{i=1}^{N_1}\mbox{Tr}(FB_{1i}\rho_{AB}
  B_{1i}^{\dag}F^{\dag})$.
On the other hand, the composition
  symmetrical channels (B) with the inter-mediate filtration produces
\begin{eqnarray} \chi_{A2}\circ F_A \circ
    \chi_{A1}\circ\chi_{B2}\circ
    F_B\circ\chi_{B1}(\rho_{AB})=\frac{1}{S}\times\nonumber\\
    \sum_{i,k=1}^{N_1}\sum_{j,l=1}^{N_2}
    B_{2i}F_BB_{1j}A_{2k}F_AA_{1l}\rho_{AB} A_{1l}^{\dag}F_A^{\dag}
    A_{2k}^{\dag}B_{1j}^{\dag}F_B^{\dag}B_{2i}^{\dag},\nonumber\\
\end{eqnarray}
where the success rate is
\begin{equation}
    S=\sum_{j,l=1}^{N_2}\mbox{Tr}(F_AA_{1l}F_BB_{1j}\rho_{AB}
    B_{1j}^{\dag}F_B^{\dag}A_{1l}^{\dag}F_A^{\dag}).
\end{equation}
The task considered here is, by the application of quantum
filters, transmit the entanglement through the composite channel
which is entanglement breaking. It means to find if the composite
channel, satisfying (\ref{cond1},\ref{cond3}) or
(\ref{cond2},\ref{cond3}),
will preserve entanglement, i.e. will
fulfil the conditions
\begin{equation} [\chi_{B2}\circ F_B\circ
\chi_{B1}(\rho_{AB})]^{T_A}<0 \end{equation} or \begin{equation}
[\chi_{A2}\circ F_A \circ \chi_{A1}\circ\chi_{B2}\circ
F_B\circ\chi_{B1}(\rho_{AB})]^{T_A}<0
\end{equation}
at a cost
of success rate of the filtration. It is impossible to solve
such the complex task analytically for all the compositions of
any channels, even for some specific channels is
sophisticated. Therefore, rather then general answer we will
analyze some physically interesting examples (for example,
previously published in Ref.~\cite{Yu07}) to demonstrate that
adaptation by single-copy filtration or even just unitary
operation can be powerful tool to stop the sudden death of
entanglement.

\section{Asymmetrical example}
The simplest example of the adaptation for the asymmetrical
configuration (A) is following. In the asymmetrical configuration, the
channel $\chi_{B1}$ ($\chi_{B2}$) transmits any qubit state either
perfectly with a probability $p_1$ ($p_2$) or, with probability
$1-p_1$ ($1-p_2$), the qubit is completely lost and another qubit in
the pure state $|0\rangle$ ($|1\rangle$) occurs in the channel. Here,
for simplicity, we use just two orthogonal states but similar analysis
can be done for two general mixed states.
It is easily to check that both the channels are entanglement
preserving for $p_1,p_2>0$. If such the channels are combined, the
maximally entangled state
$|\Psi_-\rangle=(|01\rangle-|10\rangle)/\sqrt{2}$ is transformed to
the mixture
\begin{equation}
  \rho=p_1p_2|\Psi_-\rangle\langle\Psi_-|+\frac{p_2(1-p_1)}{2}\mbox{1}\otimes
  |0\rangle\langle 0|+\frac{1-p_2}{2}\mbox{1}\otimes |1\rangle\langle
  1|,
\end{equation}
which is entangled (using partial transposition
criterion \cite{PPT}) if and only if $p_1,p_2\not=0$ satisfy
\begin{equation} p_2>\frac{1-p_1}{1-p_1+p_1^2}.
\end{equation}
Then outgoing two-qubit state has the concurrence \cite{Wootters}
\begin{equation}
C(\rho)=p_1p_2-\sqrt{(1-p_1)(1-p_2)p_2}.
\end{equation}
Otherwise the composite of these two channels is
entanglement breaking. This is the sudden death of entanglement
which apparently accompanied by the entanglement breaking. But
fortunately, if the unitary transformation making
$|0\rangle\leftrightarrow |1\rangle$
is simply applied between the
channels, then the density matrix changes to
\begin{equation}
\rho'=p_1p_2|\Phi_-\rangle\langle\Phi_-|+\frac{1-p_1p_2}{2}\mbox{1}\otimes
      |1\rangle\langle 1|,
\end{equation} where
$|\Phi_-\rangle=(|00\rangle-|11\rangle)/\sqrt{2}$, which is
always entangled for any $p_1,p_2\not=0$. All the entanglement
breaking channels vanish just simply by the unitary operation.
If two general states are considered instead of $|0\rangle$ and
$|1\rangle$, the unitary transformation depends on both the
channels. The same result can be obtained for any
maximally entangled state entering into the channels. The amount
of entanglement is simply given by the concurrence
$C'(\rho)=p_1p_2$. As a result, practically, the composite
channel is not entanglement breaking channel at all (for
$p_1p_2>0$). The entanglement can be further enhanced using
local filtering after the composite channel to approach maximal
concurrence $C''(\rho)=\sqrt{p_1p_2}$. The optimal filtering is
$|11\rangle\rightarrow \sqrt{p_1p_2}|11\rangle$ and
$|01\rangle\rightarrow\epsilon |01\rangle$, where
$\epsilon\rightarrow 0$. It is evidently a simple witness that
even unitary adaptation between the different channels can stop
the sudden death of entanglement.

\begin{figure} \centerline{\psfig{width=7cm,angle=0,file=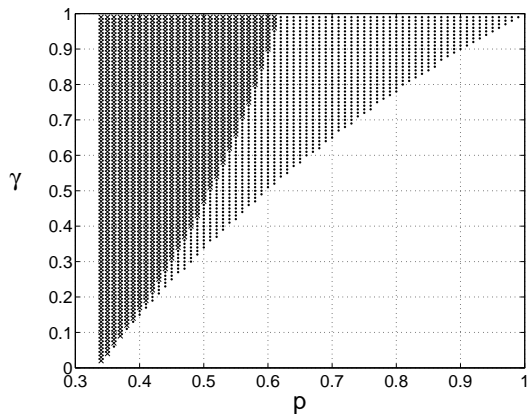}}
\caption{The quantum adaptation for the maximally entangled state passing through
symmetrical pairs of consecutive depolarizing and amplitude damping
channels. Numerical optimization has been performed in a net of the points.
Sudden death of entanglement happens for input $|\Phi_{-}\rangle$ in the
union of crosses and dots  and for input $|\Psi_{-}\rangle$
just in the area of crosses.
The entanglement breaking happens just in the area of crosses
due to possible unitary conversion between input states.
The entanglement breaking  in the area of crosses can be
undone by our quantum adaptation procedure.
 In the area where $p<1/3$, the entanglement is broken already in the
depolarizing channel, on the other hand, the white area on the right
is not interesting since the entanglement is preserving.} \end{figure}
\begin{figure} \centerline{\psfig{width=7cm,angle=0,file=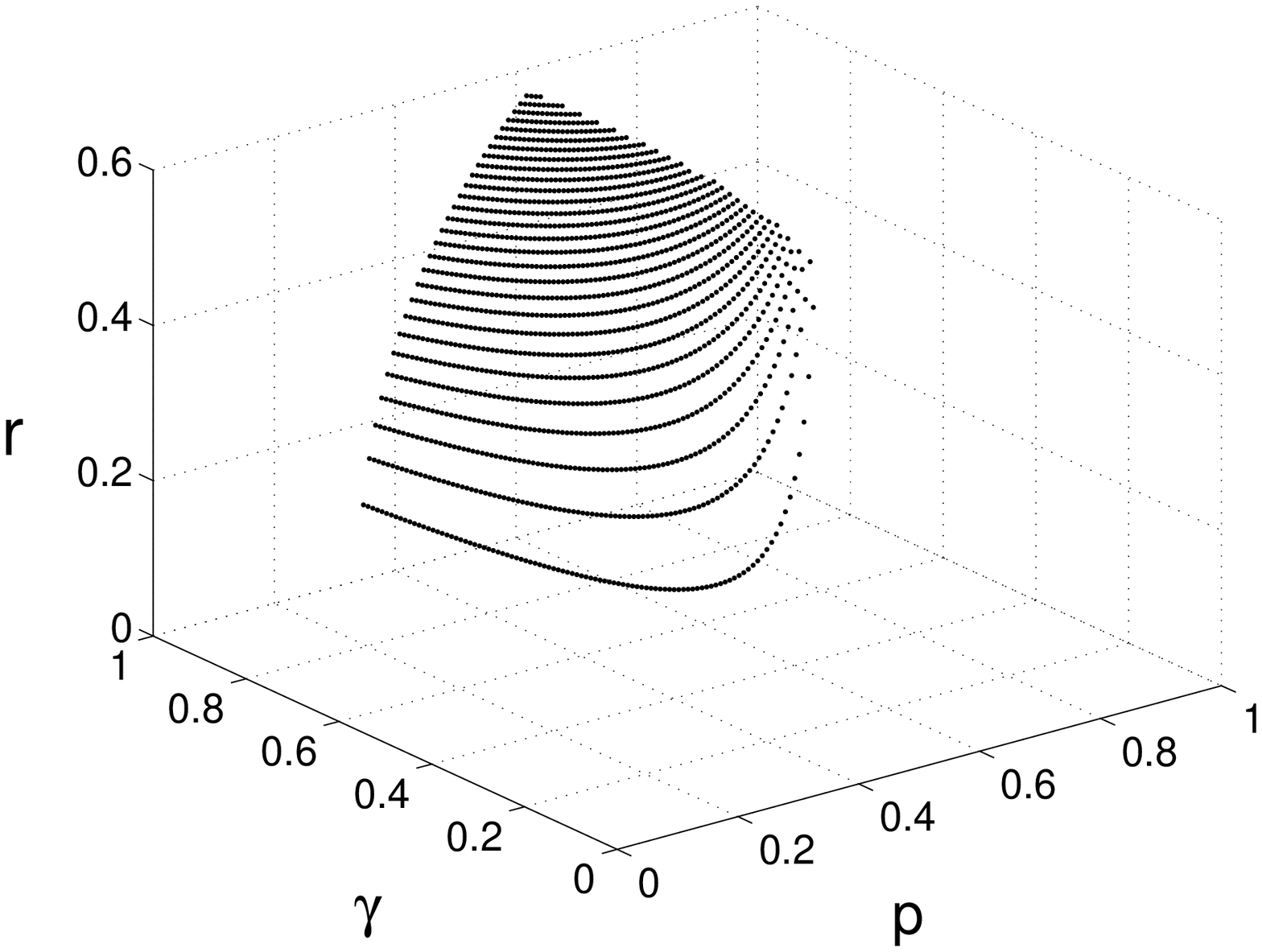}}
\centerline{\psfig{width=7cm,angle=0,file=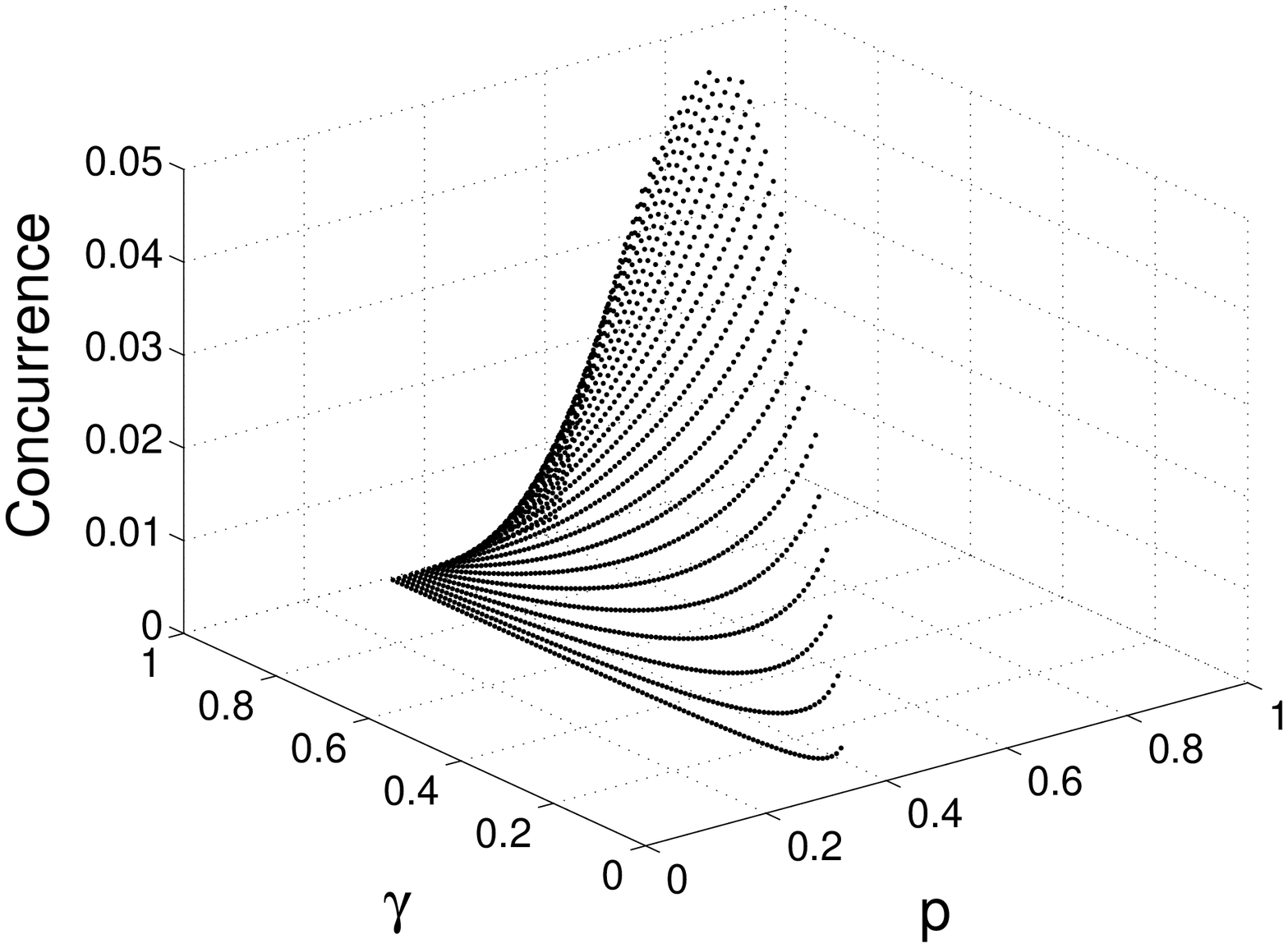}} \caption{The
plot of optimal parameter $r$ of a diagonal filter
depending on the parameters $\gamma$ and $p$ of the channels (top
figure) and the plot of concurrence depending on the parameters $\gamma$ and
$p$ of the channels (bottom figure).
Both plots are results of the quantum adaptation for the sequence of
depolarizing and amplitude damping channel. We used diagonal filters
$F_A,F_B=\mbox{diag}(1,\sqrt{r})$ for the adaptation.} \end{figure}

\section{Symmetrical example}
In the following example, we show that the sudden death of
entanglement can depend on the input maximally entangled state in
the symmetrical configuration and the adaptation by the unitary
operation is then insufficient to reduce the break of entanglement. But
using quantum filters, the break of entanglement can be
completely eliminated. Remind, if the filter is placed between the
channels to improve the transmission, it can just increase the
entanglement from the first channel at the maximum and then it can be
send through the second channel. But such the cases cannot be
understand as the adaptation, it is exactly the known single-copy
distillation \cite{Verstraete01}. The adaptation means that the filter
depends also on the parameters of the subsequent channel. Since the
filtering cannot increase the entanglement of the mixture of the Bell
states, the states with isotropic noise \cite{Werner89}
\begin{eqnarray}\label{wer}
  \rho'_1&=&p|\Psi_-\rangle\langle\Psi_-|+\frac{1-p}{4}1\otimes
  1,\nonumber\\
  \rho'_2&=&p|\Phi_-\rangle\langle\Phi_-|+\frac{1-p}{4}1\otimes 1,
\end{eqnarray} where
$|\Phi_-\rangle=(|00\rangle-|11\rangle)/\sqrt{2}$, after the first
channels are good candidates to see an effect of the adaptation. For
both the states, the entanglement is preserved if $p>1/3$. The states
(\ref{wer}) can outcome from the depolarizing channel acting on
single (or both) of the qubits.
The depolarizing channel on single qubit is represented by the set of the Kraus
operators \begin{equation}
  D_3=\sqrt{\frac{1+3p}{4}}1,\,\,\,D_i=\sqrt{\frac{1-p}{4}}\sigma_i
\end{equation} where $i=1,2,3$ and $\sigma_i$ are the Pauli matrices
\begin{eqnarray} \sigma_1=\left(\begin{array}{cc}
0 & 1  \\
1 & 0 \end{array}\right),\,\,\,\sigma_2=\left(\begin{array}{cc}
0 & -i  \\ i & 0
\end{array}\right),\,\,\,\sigma_3=\left(\begin{array}{cc}
1 & 0  \\
0 & -1 \end{array}\right). \end{eqnarray} The depolarization is common
physical decoherence process, it can arise from the random isotropic
unitary changes of the state in the channel.

Such the states (\ref{wer}) are then locally processed by the filters
$F_A,F_B$ and then entry into the identical amplitude damping
channels acting symmetrically on both the qubits. The
amplitude damping channel is non-unital channel described by the Kraus
matrices
\begin{eqnarray} A_1=\left(\begin{array}{cc}
1 & 0  \\
0 & \sqrt{1-\gamma}
\end{array}\right),\,\,\,A_2=\left(\begin{array}{cc}
0 & \sqrt{\gamma}  \\
0 & 0 \end{array}\right).
\end{eqnarray}
It often arises from a resonant interaction of qubit system with
zero-temperature reservoir characterized by a Hamiltonian $H_A=\sum_k
g_k\sigma_- a_k^{\dag}+g_k^{*}\sigma_+ a_k$, where
$\sigma_{\pm}=\frac{1}{2}(\sigma_1\pm i\sigma_2)$ are operators of the two-level
system, $a_k,a^{\dag}_k$ are the reservoir operators, $g_k$ is
coupling constant and the averaging is over the modes of reservoir,
for review
\cite{Scully}. Physically, the source of decoherence is then just the
spontaneous emission of the two-level system. This non-unital
amplitude damping channel will not break the entanglement for any
$\gamma \in (0,1)$, corresponding to finite time dynamics.
This channel has been used in Ref.~\cite{Yu07},
where non-trivial sudden death of entanglement has been recognized.

For the symmetrical configuration, the sudden death of entanglement
depends on the input maximally entangled state.
In the Fig.~2 for channel parameters in the union of crosses
and dots the sudden death of entanglement occurs for the input
state $|\Phi_{-}\rangle$ whereas for the input state $|\Psi_{-}\rangle$
sudden death of entanglement appears just in the area of
crosses. The unitary
adaptation is able to make conversion between these cases without
any reduction of the break of entanglement.
But the same effect can be obtained if the input maximally entangled
state is changed.
Therefore, in the region of dots the sudden death of entanglement
is not accompanied by the break of entanglement.
Thus unitary adaptation cannot help to stop the
break of entanglement at all, contrary to the previous example.

Interestingly, quantum filtering can help to adapt
the channels each to other. Even for this specific example,
it is complex to find the optimal filter
analytically. From this reason, the numerical genetic algorithm for
function optimization has been used \cite{genetic}.
The optimization has been performed in a net of the points and at the end, the
optimized filters have been used to check their ability to stop the
sudden death of entanglement.
There has been included, beside quantum filters, also unitary
operations into the
optimization routine.
The results of numerical calculations
are depicted in Fig.~2.
In all the numerically analyzed cases, the sudden
death of entanglement is corrected (denoted by crosses).
From the numerical optimization, it is also possible to find that a
quantum filtering is sufficient even taking both
the filters in the basis of the amplitude
damping and identical thus $F_A,F_B=\mbox{diag}(1,\sqrt{r})$ can be simply used.
From the partial transposition criterion \cite{PPT}, it is
possible to derive a sufficient condition
\begin{equation}
0<\sqrt{r}<\frac{2\sqrt{p(1+p)}-(1+p)}{\gamma(1-p)}
\end{equation}
for the quantum filters to stop the break of entanglement. The
success rate of the filtration is then $S=p(1-\sqrt{r})^2+(1+\sqrt{r})^2$. To find
optimal filter and the concurrence after
the adaptation, the numerical optimization still has to be performed.
The results are depicted on Fig.~3, where the optimal
parameter $r$ of the filter and maximal value of the concurrence are plotted as
functions of the channel parameters $\gamma$ and $p$. Evidently, in
all the cases, the filtering completely eliminates the sudden death
of entanglement caused by the symmetrical amplitude-damping channels.

\section{Conclusion}
In Conclusion, single-copy quantum adaptation of channels has been
proposed to stop non-trivial sudden death of entanglement arising in
a composite of the independent channels.
The adaptation differs from the entanglement
distillation, it rather prepares the entangled states for the
transmission through the subsequent channel. A power of both the unitary
operation and quantum filters to completely reduce the sudden death of
entanglement has been demonstrated.
The presented results have direct application for the quantum key
distribution through noisy channel and,
in an extended multi-qubit version, also for the preparation of cluster states
for quantum computing.

\medskip \noindent {\bf Acknowledgments} The research has been
supported by projects  No. MSM
6198959213 and No. LC06007 of the Czech Ministry of Education and by
the grant 202/07/J040 of GACR. R.F. also acknowledges a support by the
Alexander von Humboldt Foundation.



\begin{thebibliography}{99}

\bibitem{qc} M. Curty, M. Lewenstein and N. L\" utkenhaus, Phys. Rev.
Lett. 92, 217903 (2004).

\bibitem{qcomp} R. Raussendorf and H. J. Briegel, Phys. Rev. Lett. 86,
5188 (2001).

\bibitem{Yu04} T. Yu, and J. H. Eberly, Phys. Rev. Lett 93, 140404
(2004).

\bibitem{Yu06} T. Yu and J.H. Eberly, Opt. Comm. 264, 393 (2006).

\bibitem{Yu07} T. Yu, and J. H. Eberly, Quantum Information and
Computation 7, 459-468 (2007)

\bibitem{Davidovitch07} M.P. Almeida et al., Science 316, 579 (2007).

\bibitem{Ruskai03} M.-B. Ruskai, Rev. Math. Phys. 15, 643-662 (2003).

\bibitem{Hayashi} H. Hayashi, G. Kimura and Y. Ota, Phys. Rev. A 67,
  062109 (2003).

\bibitem{Verstraete01} M. Horodecki, P. Horodecki, R. Horodecki, Phys.
Rev. A 60, 1888 (1999); A. Kent, N. Linden, and S. Massar, Phys. Rev.
Lett. 83, 2656 (1999); F. Vertraete, J. Dehaene and B. DeMoor, Phys.
Rev. A, 010101R (2001).

\bibitem{Wang06} Z.-W. Wang, X.-F. Zhou, Y.-F. Huang, Y.-S. Zhang,
X.-F. Ren, G.-C. Guo. Phys. Rev. Lett. 96 220505 (2006)

\bibitem{Kraus} K. Kraus, States, Effect, and Operations: Fundamental
Notions in Quantum Theory (Springer- Verlag, Berlin, 1983).

\bibitem{PPT} R. Horodecki, P. Horodecki and M. Horodecki, Phys. Lett.
A 200 (5), 340 (1995); A. Peres, Phys. Rev. Lett. 77 1413 (1996).

\bibitem{Bennett96} C.H. Bennett, D.P. DiVincenzo, J.A. Smolin, and
W.K. Wootters, Phys. Rev. A 54,
3824 (1996); D. Deutsch, A. Ekert, R. Jozsa, C. Macchiavello, S.
Popescu, and A. Sanpera, Phys. Rev. Lett. 77, 2818 (1996); M.
Horodecki, P. Horodecki, and R. Horodecki, Phys. Rev. Lett. 78,
574 (1997).

\bibitem{Wootters} W.K. Wootters, Phys. Rev. Lett. 80, 2245 (1998).

\bibitem{Werner89} R. Werner, Phys. Rev. A 40,
4277 (1989).

\bibitem{Scully} M.O. Scully and M.S. Zubairy, Quantum Optics, Cambridge
University Press, Cambridge (1997).

\bibitem{genetic} See http://www.ise.ncsu.edu/mirage/GAToolBox/gaot
for the details.




\end{thebibliography}
\end{document}